%% file: madlori.tex
\documentclass[twocolumn]{aa} 
\usepackage{longtable}
\usepackage{lscape} 
\usepackage{graphicx}
\usepackage{natbib}
\usepackage{deluxetable}
\usepackage{url}
\usepackage{txfonts}
\begin{document}
   \title{A deep look into the core of young clusters\thanks{Based on observations made at the ESO La Silla and Paranal Observatory under programmes 082.C-0724, 080.D-0532(, 67.C-0042, 074.C-0084, and 074.C-0628, on observations collected at the Centro Astron\'omico Hispano Alem\'an (CAHA) at Calar Alto, operated jointly by the Max-Planck Institut f\"ur Astronomie and the Instituto de Astrof\'isica de Andaluc\'ia (CSIC) and at the Calar Alto Observatory, on observational data obtained at the Canada-France-Hawaii Telescope (CFHT) which is operated by the National Research Council of Canada, the Institut National des Sciences de l'Univers of the Centre National de la Recherche Scientifique of France,  and the University of Hawaii, on data collected at Subaru Telescope, which is operated by the National Astronomical Observatory of Japan, and on observations made with the Spitzer Space Telescope, which is operated by the Jet Propulsion Laboratory, California Institute of Technology under a contract with NASA.}}
\subtitle{II. $\lambda-$Orionis}

   \author{H. Bouy\inst{1} \thanks{Marie Curie Outgoing International Fellow MOIF-CT-2005-8389}
          \and N. Hu\'elamo\inst{2}
 	  \and D. Barrado y Navascu\'es\inst{2}
	  \and E.~L. Mart\'\i n\inst{2,3}
          \and M.~G. Petr-Gotzens\inst{4}
          \and J. Kolb\inst{4}
          \and E. Marchetti\inst{4}
          \and M. Morales-Calder\'on\inst{2}
          \and A. Bayo\inst{2}
          \and E. Artigau\inst{5}
          \and M. Hartung\inst{5}
          \and F. Marchis\inst{6}
          \and M. Tamura\inst{7}
          \and M. Sterzik\inst{8}
          \and R. K\"ohler\inst{9}
          \and V.~D. Ivanov\inst{8}
          \and D. N\"urnberger\inst{8}
          }

   \offprints{H. Bouy}

   \institute{Instituto de Astrof\'\i sica de Canarias, C/ V\'\i a L\'actea s/n, E-38200 - La Laguna, Tenerife, Spain\\
     \email{bouy@iac.es}
     \and
     Laboratorio de Astrof\'\i sica Estelar y Exoplanetas, LAEX-CAB (INTA-CSIC), PO BOX 78, E-28691, Villanueva de la Ca\~nada, Madrid, Spain\\
     \and
     University of Central Florida, Department of Physics, P.O. Box 162385, Orlando, FL 32816-2385, USA\\
     \and
     European Southern Observatory, Karl Schwartzschild Str. 2, D-85748 Garching bei M\"unchen, Germany\\
     \and
     Gemini Observatory, Southern Operations Center, Association of Universities for Research in Astronomy, Inc., Casilla 603, La Serena, Chile \\
     \and
     Astronomy Department, University of California, Berkeley, CA 94720, USA \\
     \and
     National Astronomical Observatory, 2-21-1 Osawa, Mitaka, Tokyo 181-8588, Japan\\
     \and
     European Southern Observatory, Alonso de Cordova 3107, Vitacura, Casilla 19001, Santiago 19, Chile \\
     \and
     Landessternwarte, Zentrum f\"ur Astronomie der Universit\"at Heidelberg, K\"onigstuhl, D-69117 Heidelberg, Germany
   }

   \date{Received ; accepted 01-07-2009}

 
  \abstract
   {Over the past years, the $\lambda-$Orionis cluster has been a prime location for the study of young very low mass stars, substellar and isolated planetary mass objects and the determination of the initial mass function and other properties of low mass cluster members.}
   {In the continuity of our previous studies of young associations cores, we search for ultracool members and new multiple systems within the central 5\farcm3 ($\approx$0.6~pc) of the cluster. }
   {We obtained deep seeing limited J, Ks-band images of the 5\farcm3 central part of the cluster with NTT/SofI and H-band images with CAHA/Omega2000. These images were complemented by multi-conjugate adaptive optics (MCAO) H and Ks images of the 1\farcm5 central region of the $\lambda-$Orionis cluster obtained with the prototype MCAO facility MAD at the VLT. The direct vicinity of the massive $\lambda-$Ori O8III-star was probed using NACO/SDI at the VLT. Finally, we also retrieved Spitzer IRAC images of the same area and used archival Subaru Suprime-Cam and CFHT CFHT12K $i$-band images. }
   {We report the detection of 9 new member candidates selected from optical and near-IR color-color and color-magnitude diagrams and 7 previously known members. The high spatial resolution images resolve 3 new visual multiple systems. Two of them are most likely not members of the association. The third one is made of a brown dwarf candidate companion to the F8V star HD~36861C. The simultaneous differential images allow us to rule out the presence of visual companions more massive than M$>$0.07~M$_{\sun}$ in the range 1--2.5\arcsec, and M$>$0.25~M$_{\sun}$ in the range 0\farcs5--2.5\arcsec. }
   {}

   \keywords{Stars: Evolution, Stars: formation, Stars: general, Stars: low mass, brown dwarfs, Techniques: High Angular Resolution, Stars: visual binaries}

   \maketitle
%

\section{Introduction}
The Orion's Head, when inspected visually,  is an unconspicuous area of a mythological giant. However, when observed in the X-ray or with a narrow band filter, such as H$\alpha$, or at longer wavelengths (mid- and far-IR, or CO lines), it displays a very large circular structure about 10\degr\, across the sky. Right at the center of this bubble is the O8III double star $\lambda$~Ori, which dominates the Collinder~69 cluster located at 325--450~pc \citep{1997A&A...323L..49P}. The first comprehensive studies of this association were carried out at optical wavelengths by \citet{1999AJ....118.2409D, 2001AJ....121.2124D, 2002AJ....123..387D}, who studied the stellar population and established that Collinder~69 is about 5~Myr. Later on, \citet{2004ApJ...610.1064B, 2005AN....326..981B, 2007ApJ...664..481B,2007A&A...468L...5B} collected additional optical, near- and mid-IR photometry, as well as low-resolution spectroscopy, to probe  well inside the substellar domain. Among other results, they derived an initial mass function (IMF) covering the range from the most massive member of the cluster ($\lambda$~Ori itself, about 20-30~M$_{\sun}$) down to 0.020~M$_{\sun}$, studied the accretion, and derived the ratio between Class~II and Class~III objects (Classical TTauri versus disk-less objects). \citet{2007A&A...468L...5B} reported the detection of a substellar binary candidate and \citet{2008A&A...488..167S} studied the frequency of low mass (0.2--1~M$_{\sun}$) spectroscopic multiple systems of the associations. .

\begin{figure*}
   \centering
   \includegraphics[width=\textwidth]{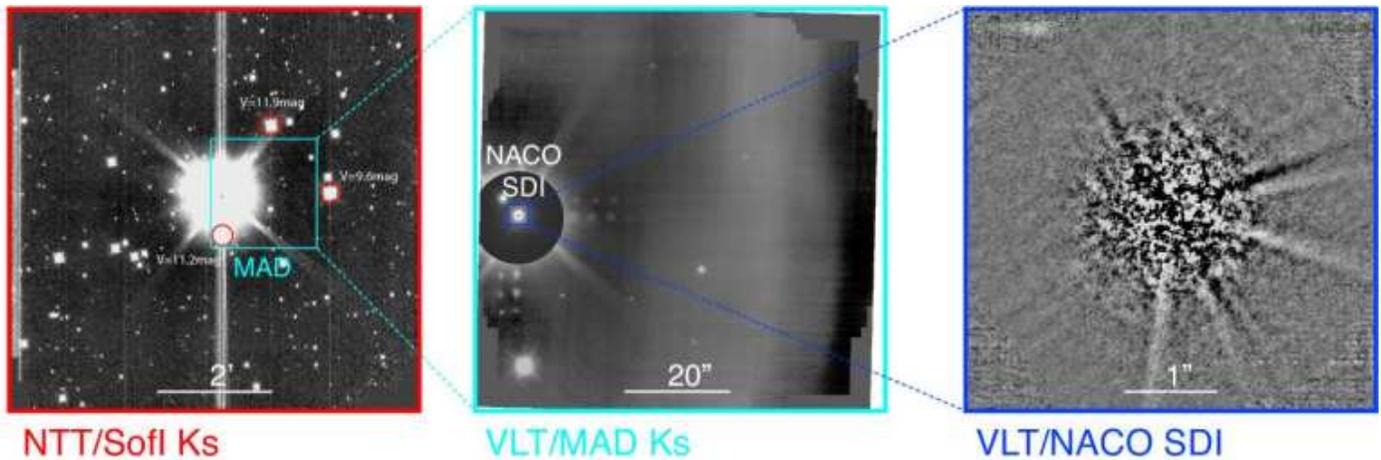}
      \caption{Overview of the multiscale observations. The MAD (cyan) field of view is represented over the SofI Ks band image (left panel, red). The 3 reference stars used for wavefront sensing with MAD are indicated with red circles. Successive zooms performed with the different instruments show in more details the spatial scales probed by MAD (center panel) and NACO/SDI (right panel, blue). The cut levels around the bright O star in the MAD image have been set differently so that the reader can appreciate the dynamic range of the image. The scales are represented. North is up and east is left.}
         \label{multiscale}
   \end{figure*}

In the continuity of our study of young associations cores \citep[see Paper~I and ][]{2008A&A...477..681B}, we present an analysis of the 5\farcm3$\times$5\farcm3 central part of the $\lambda-$Orionis cluster. The presence of the bright massive OB pair in the center of the cluster complicates the observations, as the heavy saturation that they produce on detectors limit the sensitivity of the images and prevents the detection of faint very low mass members in their vicinity. To circumvent this problem, we combine datasets of several instruments of increasing resolving powers and sensitivities. Deep wide field near-infrared (near-IR) images covering the inner 5\farcm3$\times$5\farcm3 of the cluster were obtained with SofI \citep{Sofi} at the NTT  in La Silla and with Omega2000 \citep{1998SPIE.3354..825B} at the 3.5~m telescope of the Calar Alto Observatory. MCAO images of the central 1\farcm5 region were obtained using MAD \citep{2007Msngr.129....8M} at the VLT in Paranal. Deep CFHT12K \citep{2000SPIE.4008.1010C} and Subaru Suprime-Cam \citep{2002PASJ...54..833M} $i$-band images from \citet{2004ApJ...610.1064B} and \citet{2008PhDT_MariaMorales} as well as archival {\it Spitzer} IRAC images \citep[program 37, P.I. Fazio][]{2007ApJ...664..481B} overlapping with the near-infrared images are also re-analyzed. Finally, the direct vicinity of the central O8III star $\lambda-$Ori was probed using NACO and its simultaneous differential imaging (SDI) mode at the VLT in Paranal \citep{2003SPIE.4841..944L, 2003SPIE.4839..140R}. Figure~\ref{multiscale} gives an overview of the observations.

\section{Wide field imaging}

\subsection{NTT/SofI J and Ks band images} 
Deep J-band images were obtained on 2008 December 20 with SofI at the NTT as part of programme 082.C-0724 (P.I. Morales Calder\'on). A set of 40 dithered exposures of 12$\times$5s (\verb|NDIT|$\times$\verb|DIT|) was obtained, leading to a total on-source exposure time of 40~min. The images were processed following standard procedures using the \emph{Eclipse} reduction package \citep{1997Msngr..87...19D} and the final mosaic covers 5\farcm5$\times$5\farcm4 centred on $\alpha=$05$h$35\arcmin 08\farcs6 and $\delta=$ +09\degr 55\arcmin 52\farcs.2 (J2000). A photometric standard star \citep[$\lbrack$PMK98$\rbrack$ 9108, ][]{1998AJ....116.2475P} was observed immediately before the target. The night was clear, and the ambient conditions were good with a seeing in the J-band of 0\farcs8 as measured on the final image. 

The cluster has been observed in the Ks band with SofI at the NTT on 2001 December 04 (P.I. Testi, Programme 67.C-0042). A set of 15 dithered images of 12$\times$5~s (\verb|NDIT|$\times$\verb|DIT|) was obtained that night. We retrieved the data and the corresponding calibration frames and processed them following the same procedures. The seeing in the Ks-band measured on the image was 0\farcs9, and the final mosaic covers 4\farcm9$\times$4\farcm8 centred on $\alpha=$05$h$35\arcmin05\farcs7 and $\delta=$ +09\degr55\arcmin57\farcs3 (J2000). 

\begin{figure}
   \centering
   \includegraphics[width=0.45\textwidth]{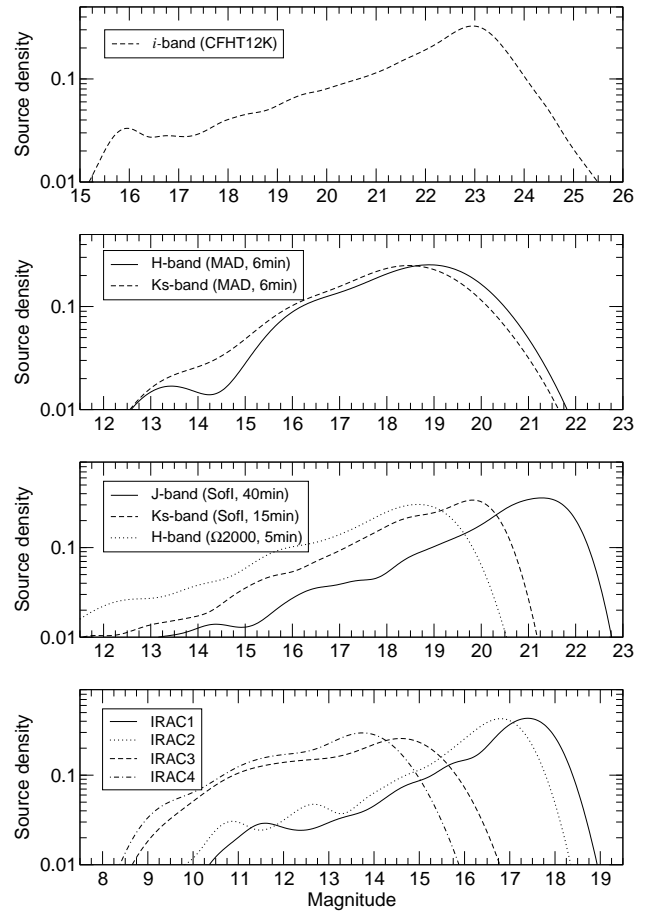}
      \caption{Distribution of magnitude of the sources detected in the CFHT12K, MAD, SofI, $\Omega$2000 and IRAC images, computed using a kernel density estimate \citep{kernel_estimate_2d}. The drop at faint magnitudes gives a good approximation of the completeness limit of the images.\label{limcomp}}
   \end{figure}

PSF photometry of all the sources brighter than the 3-$\sigma$ noise of the local background was extracted using \citet{2000SPIE.4007..879D} Starfinder code. The counts were translated into magnitudes using the zeropoint measured using the photometric standard star (ZP=23.00$\pm$0.06~mag) in the case of the J-band observations, and using eight well behaved (quality flag \verb|A|) unresolved 2MASS sources in the case of the Ks-band archival data for which no standard star was available (ZP=23.99$\pm$0.09~mag). The faintest sources detected in the images\footnote{far from the bright central stars} reach J$\approx$22.5~mag and Ks$\approx$21~mag and Fig.~\ref{limcomp} shows that the limit of completeness\footnotemark[1] reaches J$\approx$21~mag and Ks$\approx$20~mag. On the bright side, the detector non-linearity reaches about 3\% at 14000~A.D.U., corresponding to J=11.6~mag and Ks=12.4~mag. The final processed mosaics are made available upon request from the authors of this article. 

\subsection{Omega2000 H band images}
Deep near-IR H-band images were obtained with the Calar Alto 3.5m telescope and its Omega2000 camera in October 2005 \citep[15\farcm36$\times$15\farcm36, P.I. Barrado y Navascu\'es; ][]{2007ApJ...664..481B}. Five dithered exposures of 60~s each were taken. The photometric calibration was derived from well behaved (quality flag \verb|A|) unresolved 2MASS counterparts. The final uncertainties are estimated to be 0.05~mag. The present study focuses only on the overlap between the Omega2000 and SofI images. Figure~\ref{limcomp} shows that the final image is complete\footnotemark[1] up to H$\approx$19~mag. The faintest source detected has H$\approx$21~mag. 

\subsection{CFHT12K $i$-band images}
Deep $i$-band images were obtained at the Canada France Hawaii Telescope (CFHT) with the CFHT12K mosaic camera on 1999 September 29 and 30. The observations are described in details in \citet{2004ApJ...610.1064B}. We extracted the PSF photometry using \citet{2000SPIE.4007..879D} Starfinder code and used the photometric zero point (ZP=26.121$\pm$0.012) provided by the CFHT team as part of the calibration plan for service mode observations. Figure~\ref{limcomp} shows that the CFHT12K catalogue is complete\footnotemark[1] up to $i\approx$23~mag, and the faintest source has $i\approx$25~mag. 

\subsection{Suprime-Cam $i$-band images}
Deep $i$-band images were obtained at the Subaru telescope with the Suprime-Cam camera in the $i$-band on 2006 November 20. The observations and data processing are described in \citep{2008PhDT_MariaMorales}. A set of 45 dithered images of 135~s each was acquired, leading to a total on-source exposure time of 6120~s. Even though the Suprime-Cam mosaic goes significantly deeper than the CFHT12K one, the OB stars saturation and halo in the region of interest of our survey are stronger. We therefore use the Suprime-Cam image only to extract $i$-band PSF photometry of the sources within the field of the high resolution MCAO images (see below) without CFHT12K counterpart. The photometric zeropoint  (ZP=26.56$\pm$0.017) is estimated using the CFHT12K photometry.

\section{High resolution MCAO images}

\subsection{Observations}
A region of ~1\farcm3$\times$1\farcm3\, centered on the $\lambda$--Orionis cluster was observed with MAD in November 2007. MAD is a prototype instrument performing wide field-of-view, real-time correction for atmospheric turbulence \citep{2006SPIE.6272E..21M}. A detailed description of the instrument and data processing is given in Paper~I and \citet{2007Msngr.129....8M}. Figure~\ref{multiscale} gives an overview of the pointing and of the guide stars used for wavefront sensing. The guide stars geometrical distribution is relatively symmetric but the luminosities of two of the reference stars were close to the wavefront sensor limit, forcing us to use a slow correction rate.

\begin{figure*}
   \centering
   \includegraphics[width=0.9\textwidth]{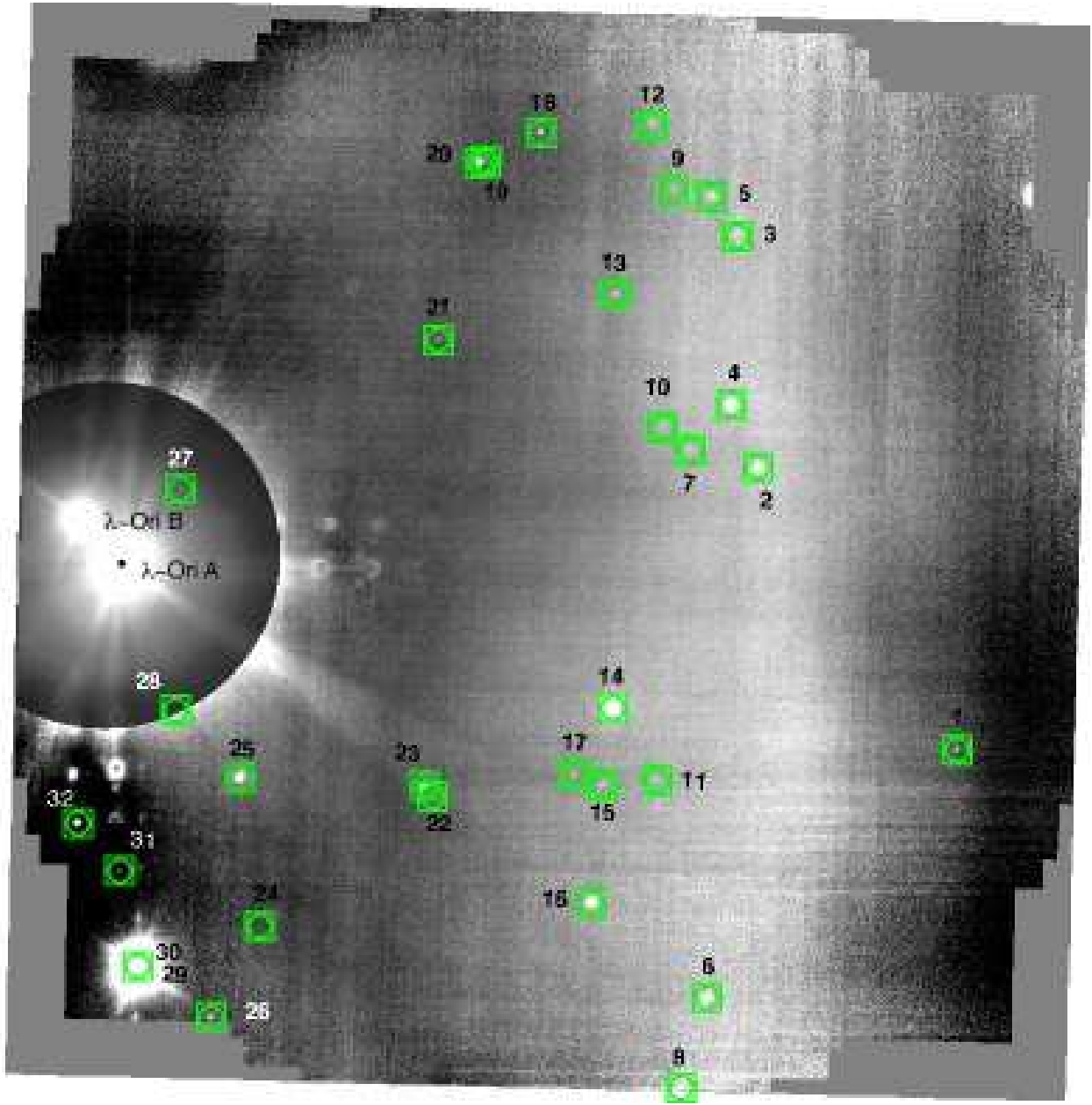}
      \caption{MAD H-band finding chart of all the sources.  All LOri-MAD candidates are indicated with a box-circle and their number. The levels around the bright binary $\lambda-$Ori~AB have been stretched differently to enhance the contrast. The light leak is clearly seen as a broad bright vertical stripe on the west half of the image. The field of view is 1\farcm3$\times$1\farcm3 wide. North is up and east is left.}
         \label{madlori}
   \end{figure*}

A set of \verb|NEXP|=30 dithered images of 30$\times$0.8~s (\verb|NDIT|$\times$\verb|DIT|) each was obtained in H and Ks, leading to a total exposure time of 12~min in each band. The ambient conditions\footnote{At the zenith and in the visible.} during the observations reported by the ESO Ambient Conditions Database are given in Table \ref{ambient}. The coherence time of the atmospheric variations was average ($\tau_{0}\approx$3~ms\footnote{Median coherence time of the atmosphere in Paranal between 1999 and 2003: 3.3~ms (source ESO)}) during both the H and Ks observations but the seeing was significantly better during the H-band observations. The final processed mosaics are made available upon request from the authors of this article. Figure~\ref{madlori} shows the final H-band mosaic.

\begin{center}
\begin{table}
\caption{Ambient conditions at the zenith and in the visible during the MAD observations}
\label{ambient}
\begin{tabular}{lcccc}\hline\hline
Filter        & date                 & airmass       & seeing     & $\tau_{0}$      \\
              & [UT]                 &               & [\arcsec]  & [ms]           \\
\hline
Ks            & 2007-11-30 07:21 & 1.35$\pm$0.03 & 1\farcs2$\pm$0\farcs2   & 3.2$\pm$0.6  \\ 
H             & 2007-12-01 06:48 & 1.28$\pm$0.02 & 0\farcs77$\pm$0\farcs10 & 2.6$\pm$0.3 \\
\hline
\end{tabular}
\end{table}
\end{center}

\subsection{MCAO performances}
\begin{figure}
   \centering
   \includegraphics[width=0.45\textwidth]{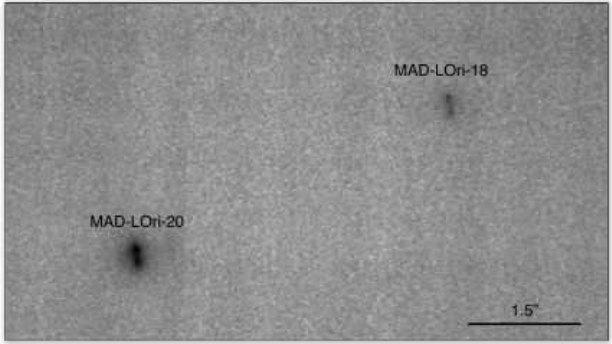}
      \caption{Elongated and doubled PSF in the MAD image (Ks). The scale is indicated. North is up and east is left.}
         \label{doblitos}
   \end{figure}

As expected the quality of the AO correction follows closely the geometry of the 3 reference stars. A number of sources appear to be artificially elongated or doubled, as shown in Fig.~\ref{doblitos}. These local artefacts appear in the individual images, and might be due to the re-centering of the guide stars during the observing sequence leading to a field-of-view warping with unpredictable shape. The 9 affected sources are indicated in Table~\ref{table_mad}. Their properties (duplicity and luminosities) are less reliable than for the other objects.

The Strehl ratio in the Ks band ranges from 23\% to 27\% (not including objects suffering from doubling/elongation). The under-sampling of the point spread function (PSF) in the H band prevents us to compute meaningful Strehl ratios, but the performances are expected to be similar. 

\subsection{Photometry}
The strategy adopted to extract the photometry is described in Paper~I. Briefly, we performed aperture photometry except in a few cases of close multiple systems or when the source was located in the halo of a bright massive star (LOri-MAD-29, -27, -28). In these latter cases we extracted the photometry using nearby isolated stars as reference PSF. The aperture photometry was performed using standard routines with the \emph{daophot} package within IRAF\footnote{IRAF is distributed by the National Optical Astronomy Observatory, which is operated by the Association of Universities for Research in Astronomy (AURA) under cooperative agreement with the National Science Foundation.}, using an aperture of 11~pixels (0\farcs31), and a sky annulus between 12--16~pixels (0\farcs34--0\farcs45). Such an aperture is well suited even for the abnormally elongated and doubled objects as it includes most of their flux. Ten well-behaved isolated and unresolved sources from Omega2000 H-band catalog were used to derive the zeropoints given in Table \ref{zp}. The Ks-band zeropoint was computed using 8 clean and unresolved matches found in the SofI Ks image. One pair of objects (LOri-MAD-20 \& -19) required a specific photometric extraction. The two objects form a visual pair separated by only 0\farcs515, so that the 11~pix aperture (0\farcs308) was not well suited. Instead, we used a 9~pix aperture (0\farcs252)  and a sky annulus between 40-44 pixels to avoid contamination by the visual companion. The instrumental zeropoints corresponding to this aperture extraction are 24.449$\pm$0.025~mag in H and 23.615$\pm$0.061~mag in Ks. A couple of objects are not detected in the Ks band and one object is detected in the Ks-band only. An upper limit on their luminosity was derived by adding artificial stars of decreasing luminosity at the expected position of the source until the 3-$\sigma$ detection algorithm misses it. The photometry is presented in Table~\ref{table_mad}.

Sources brighter than $\approx$8~mag are saturated or above the detector linearity limit. We detect sources as faint as Ks=20.3~mag and H=20.7~mag. Because of the halo of the bright OB binary, and because of some light leak in the camera, the limit of sensitivity is not homogeneous over the whole field-of-view. The light leak appears as a broad bright stripe on the west half of the images, as illustrated in Fig.~\ref{madlori}. It is present and variable in all individual images and was not efficiently removed by the sky-subtraction process. The stronger background due to this leak varies only slowly and smoothly over the image, and should not affect the aperture photometry. In the most affected areas, the background level {\bf (after sky subtraction)} can be five times higher than in unaffected areas.

\begin{center}
\begin{table}
\caption{Instrumental zeropoints for the MAD observations}
\label{zp}
\centering
\begin{tabular}{lll}\hline\hline
Method    &   Filter    & Zeropoint [mag]   \\
\hline
Aperture  & H         & 24.45$\pm$0.05 \\
Aperture  & Ks         & 24.26$\pm$0.03 \\
PSF       & H         & 25.06$\pm$0.09 \\
PSF       & Ks         & 24.30$\pm$0.09 \\
\hline
\end{tabular}
\end{table}
\end{center}

\subsection{Nature of the detections}
We compare the MAD and SofI detections to rule out the possibility that some of the MAD sources are artefacts (such as e.g bad pixels, remnants, ghosts, cosmic ray events, etc.). Only two objects detected in the MAD images do not have a SofI counterpart (LOri-MAD-19 and 24), ruling out the possibility of false detections for the other 30 objects. They are not detected in the SofI images either because of the lack of sensitivity (LOri-MAD-24) or spatial resolution (LOri-MAD-19). LOri-MAD-30 is detected in the MAD Ks-band only. Until confirmed with new images, these three sources should be considered with caution. All but 5 MAD sources have counterparts in the SofI J-band image. Four of them are either too close to the bright OB pair (LOri-MAD-27 and LOri-MAD-28, with 3-$\sigma$ limits of J$>$15.1~mag and J$>$17.6~mag respectively) or blended with a nearby companion (LOri-MAD-19 and LOri-MAD-30). The fifth one, LOri-MAD-24 is not detected with a 3-$\sigma$ limit of J$>$20.1~mag. Finally, the bright LOri-MAD-29 is detected but saturated, and we report its 2MASS J-band photometry assuming that the contribution of its close visual companion LOri-MAD-29 is negligible.

\section{{\it Spitzer} IRAC images}

The $\lambda-$Orionis cluster was observed with {\it Spitzer} IRAC on 2004 October 11 in the course of program 37 \citep[P.I. Fazio,][]{2007ApJ...664..481B}. Program 37 was executed in High Dynamic Range (HDR) mode providing equal numbers of consecutive short (0.6~s) and long (12~s) exposures. We retrieved the calibrated, individual IRAC BCD (Basic Calibrated Data) images and stacked them following the procedures recommended by the Spitzer Science Center (SSC) with the MOPEX software package and the relevant calibration files. The short exposure mosaics allowed us to extract the photometry of bright sources otherwise saturated in the long exposure mosaics. Uncertainties were tentatively estimated from the Poisson noise weighted by the coverage maps of the mosaics. The presence of the bright and asymmetric halo, diffraction spikes and ghosts around the massive stars make it difficult to estimate reliable uncertainties in the central part. In the rest of the image, uncertainties are in general dominated by the zeropoint uncertainties  (4.1~Jy, 2.6~Jy, 1.7~Jy and 0.94~Jy for channel 1 to 4 respectively, {\bf corresponding to 0.015, 0.014, 0.015 and 0.015~mag}, as given by the {\it Spitzer} Science Center\footnote{\url{http://ssc.spitzer.caltech.edu/irac/calib/}). The final error (combining both measurement and zeropoint uncertainties) ranges between 0.01~mag and 0.08~mag for most sources.}.

\section{High resolution NACO/SDI images}

The O8III star $\lambda-$Ori was observed with NACO/SDI at the VLT on 17 November 2007 and 12 December 2007 as part of program 080.D-0532 (P.I. Bouy). The ambient conditions during the second observations were worse and the AO performance and final sensitivity of the images were degraded. For the rest of our analysis, we refer only to the first dataset. The target was used as reference for the wavefront sensing. CONICA's SDI mode was designed to detect methane rich companion (T-dwarfs) around bright stars. Four images are obtained simultaneously through 3 narrow band filters, providing 4 images with the (almost) exact same quasi-static speckle pattern\footnote{within a spatial scaling factor equal to the ratio of the filters central wavelengths}. Two images are taken outside the 1.625$\mu$m methane feature (at 1.575~$\mu$m and 1.600$\mu$m) and two images are taken inside the methane feature (both at 1.625$\mu$), allowing efficient removal of the PSF quasi-static aberrations in a wavelength range were the contrast between the hot primary and a potential methane companion is optimum. All filters have a FWHM of 25~nm. A detailed description of the technique and of the instrument is given in e.g \citet{2007ApJS..173..143B}. Four series of images were obtained at 2 position angles (0\degr\, and 33\degr) following the sequence 0\degr--33\degr--33\degr--0\degr. Each series was made of 8 on-source dithered exposures and 2 off-sky images of 30$\times$2~s (\verb|NDIT|$\times$\verb|DIT|). The individual images were dark subtracted and flat-fielded. Each series of two off-sky images were average combined and subtracted to the corresponding individual images. The exposure time was chosen to slightly saturate the inner $\approx$3 pixels of the PSF core, enhancing the sensitivity at large radii.

   \begin{figure*}
   \centering
   \includegraphics[width=\textwidth]{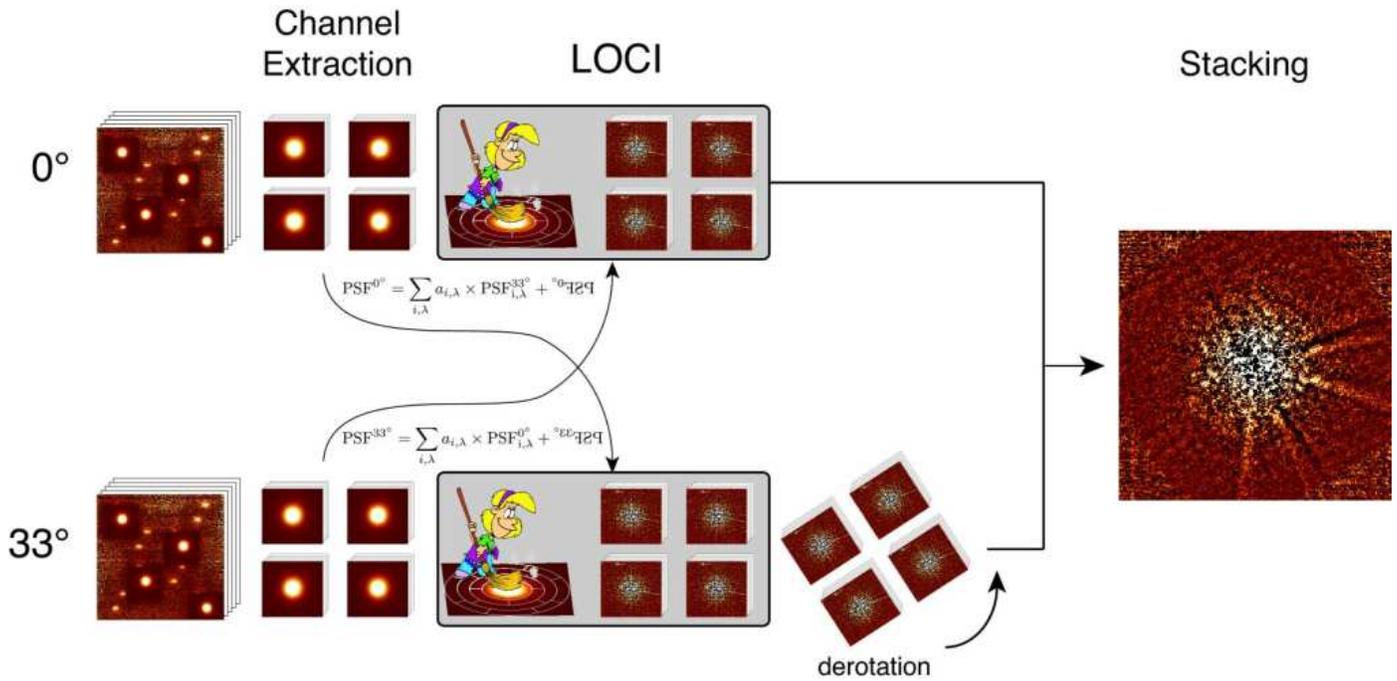}
      \caption{Cartoon summarizing the different steps in the processing of the SDI data. First the individual channels are extracted for each individual image and each rotation angle. Then for each individual image, a locally optimized PSF is computed using the LOCI algorithm and subtracted. This optimized PSF is made of a linear combination of all the rotated individual images in all 4 channels plus the reflected copy of the individual image that is fitted. The 33\degr\, output images are then de-rotated, and median combined with all the 0\degr\, images to produce the final image. }
         \label{sdi_loci}
   \end{figure*}

The difference of magnitude between a methane companion (H$\approx$20.2~mag at 5~Myr and 400~pc according to the DUSTY models and assuming that the L/T transition occurs around 1\,500~K) and the massive $\lambda-$Ori (H=3.7~mag) is out of reach of NACO/SDI. Since we do not expect to be sensitive to methane companions, we adopt a strategy different from the "classic" SDI approach but nevertheless taking advantage of the stability of the quasi-static speckle pattern of the 4 SDI channels. Figure~\ref{sdi_loci} gives an overview of the different steps involved in the processing. Each channel of each individual image is processed independently. For each of them, the stellar PSF speckles are attenuated by subtracting an optimized PSF obtained using the locally optimized combination of images (LOCI) algorithm detailed in \citet{2007ApJ...660..770L}. Briefly, this algorithm divides an image into subsections. For each subsection independently, an optimized estimate of the PSF is then built using the linear combination of a set of reference PSFs that minimizes the noise. In our case, the set of reference PSFs is made of all the \emph{rotated} images in all 4 channels plus the reflected copy of the individual image that is fitted, included to efficiently remove the symmetric pattern of the PSF. The output individual images are then registered to a common field angle and median-combined to produce the final image. The remaining speckles add up incoherently in the final image, while any companion would add up coherently. Figure~\ref{multiscale} shows the final image (4\farcs3$\times$4\farcs3). 

\begin{figure}
   \centering
   \includegraphics[width=0.45\textwidth]{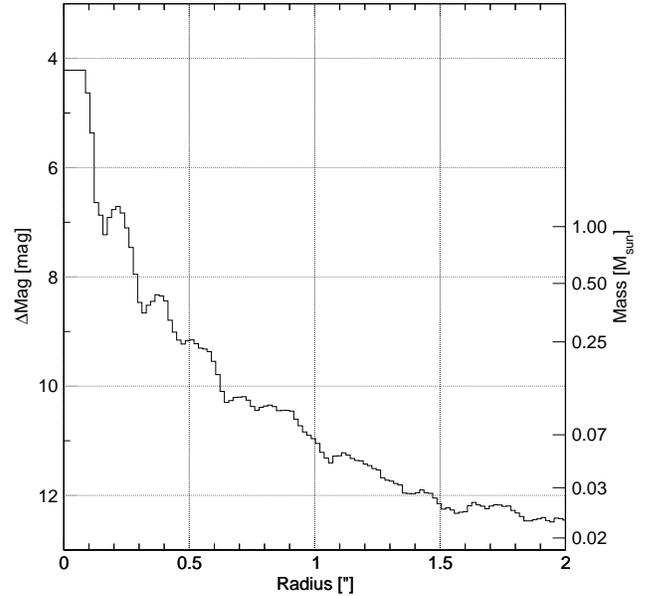}
      \caption{3$-\sigma$ limit of sensitivity curve obtained with NACO SDI and the LOCI algorithm. The curve was computed from the 3-$\sigma$ standard deviation of the radial profile. The right scale gives the mass corresponding to the same H-band difference of magnitude according to the NextGen models {\bf of \citet{1998A&A...337..403B}}. It is only indicative as the combination of the three SDI filters is close to but different from an H-band filter. Some structures are visible at $\approx$0\farcs2 and 0\farcs4, and most likely correspond to electronic and optical ghosts present in bright objects observations (see NACO's User manual). \label{sdi_contrast}}
   \end{figure}

Figure~\ref{sdi_contrast} shows that the NACO/SDI images were sensitive to brown dwarf companion as close at $\approx$1\farcs. The contrast achieved would have allowed us to detect any brown dwarf companion with a mass greater than 0.03~M$_{\sun}$ between 1\farcs5--2.5\arcsec. We do not report any such companion. This result is consistent with the results obtained for massive members of the Sco~OB2 association \citep{2005A&A...430..137K} as well as with the latest numerical simulations of star formation predicting that the lightest objects should either merge with the massive primary or be ejected during the early stages of evolution of the parent proto-stellar cluster \citep{2009MNRAS.392..590B,2009MNRAS.392.1363B}.

\section{Cluster member selection}
In the following sections and for consistency with the previous study of \citet{2004ApJ...610.1064B} and \citet{2007ApJ...664..481B}, we assume an age of 5Myr, a distance of 400~pc and a reddening of $E(B-V)=$0.12 \citep{1994ApJS...93..211D}.

\subsection{SofI-Omega2000-CFHT12K-Spuprime-Cam catalogue}

\begin{figure}
   \centering
   \includegraphics[width=0.45\textwidth]{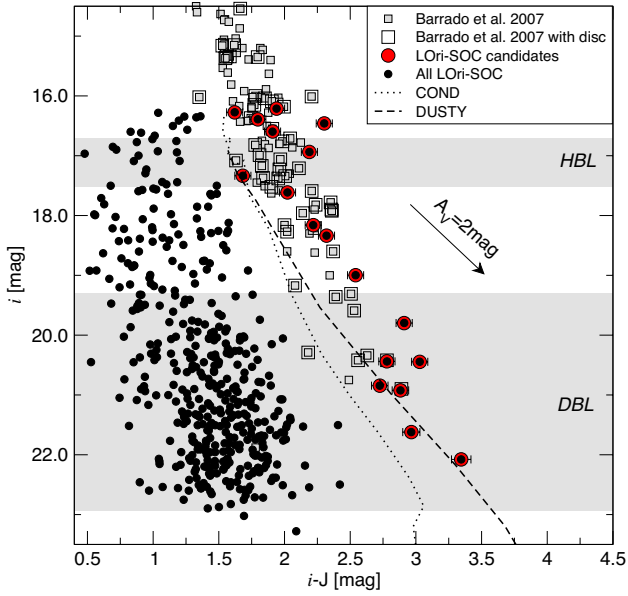}
      \caption{$i$ vs $i$-J color-magnitude diagram of the LOri-SOC sample (black dots). The DUSTY and COND isochrones at 5~Myr and 400~pc shifted blueward by 0.1~mag are represented with a dashed and dotted line respectively. Confirmed members from \citet{2007ApJ...664..481B} are represented with grey squares. Objects of their sample with mid-IR excess associated to a disc are overplotted as large open squares. Their $I_{\rm c}$ luminosities have been transformed to Sloan photometry using the empirical transformation given in \citet{2006A&A...460..339J}. The hydrogen (HBL) and deuterium burning limits (DBL) according to the DUSTY models for ages between 1--5~Myr and distances in the range 325--450~pc are  represented with grey shaded areas. Objects selected as candidate members are overplotted with a red dot. An A$_{\rm V}$=2~mag reddening vector is represented.  \label{i_ij_soc}}
   \end{figure}

\begin{figure}
   \centering
   \includegraphics[width=0.45\textwidth]{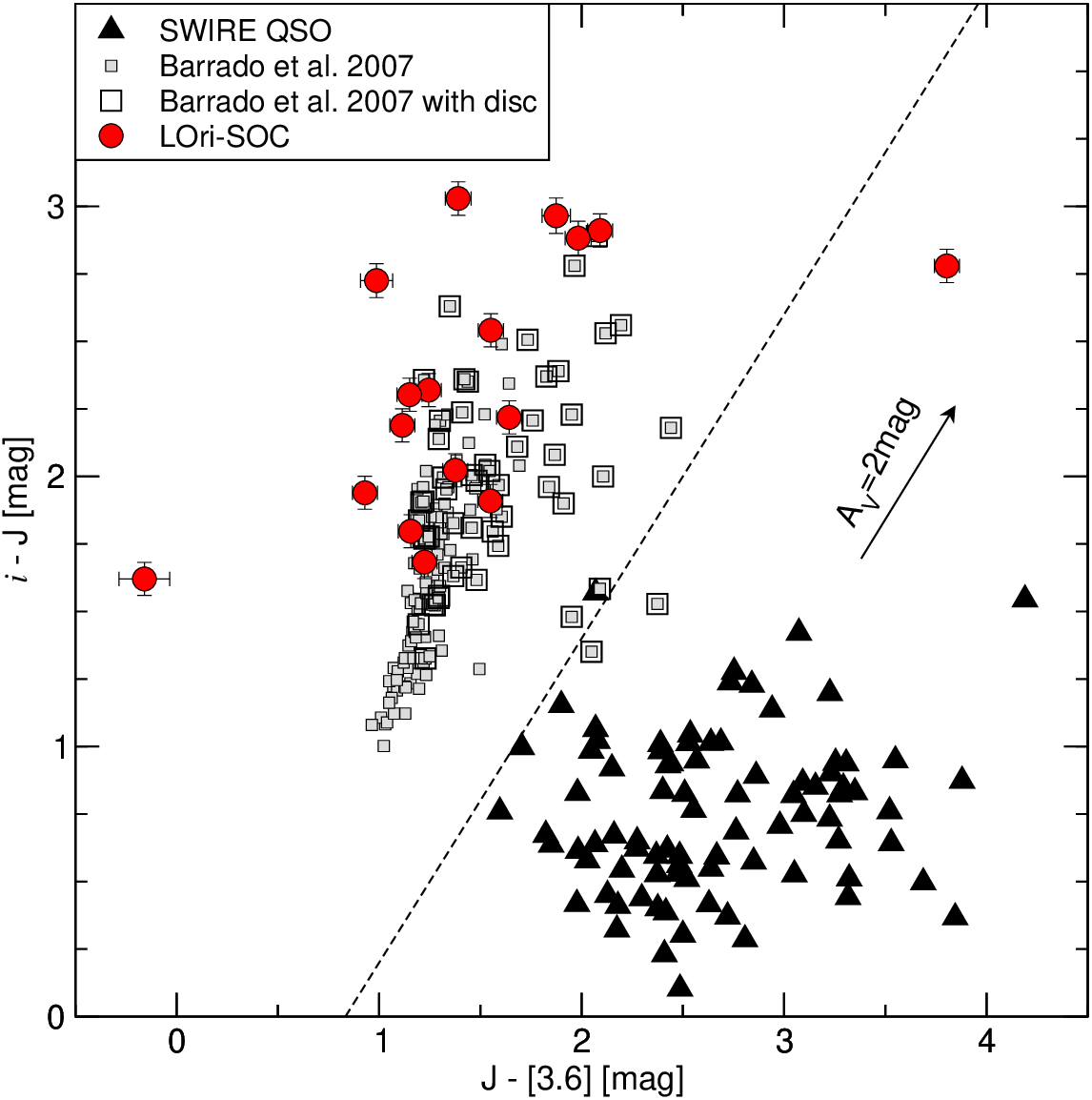}
      \caption{$i$-J vs J-[3.6] color-color diagram of confirmed $\lambda-$Ori very low mass members from \citet{2007ApJ...664..481B} (grey squares for the whole sample and an additional large open square for objects of their sample with mid-IR excess associated to a disc) and quasi stellar objects from the SWIRE survey \citep{2008MNRAS.386.1252H}. The two populations occupy very distinct areas of the diagram, which can be used to efficiently discard QSO. The LOri-SOC candidate members are overplotted as large red dots. The dashed line represent the selection criterion given in Equations~\ref{sel1}. An A$_{\rm V}$=2~mag reddening vector is represented. \label{ij_j36_soc}}
   \end{figure}

The completeness of our final catalogue for the central inner 5\arcmin\, of the cluster reaches $i$=23~mag, J=21.5~mag, H=19~mag, Ks=20~mag and [3.6]=17.5~mag (Fig.~\ref{limcomp}), therefore well below the deuterium burning limit at the age (5~Myr) and distance (400~pc) of the cluster \citep[predicted at $i$=22.14~mag, H=18.17~mag, Ks=17.64~mag for the DUSTY models][]{2000ApJ...542..464C}. We use the two deepest bands ($i$ and J) to select candidate members in a $i$ vs $i$-J diagram (Fig.~\ref{i_ij_soc} of all sources of the SofI-$\Omega$2000-CFHT12K catalogue (hereafter LOri-SOC catalogue). A total of 18 sources have $i$-band luminosities and $i$-J color consistent with the cluster isochrones and we select them as candidate members. The contamination by extragalactic sources can be high, as illustrated by the recent survey of \citet{2007ApJ...662.1067H} in the nearby $\sigma-$Orionis cluster. In exploring filtering methods to identify and reject extragalactic contaminants, we have developed a strategy taking advantage of the multiple wavelengths available in our catalogue. All 18 candidate members but one have a H, Ks and IRAC 3.6~$\mu$m counterpart. Figure~\ref{ij_j36_soc} shows a $i$-J vs J-[3.6] color-color diagram of the confirmed very low mass members of the association presented in \citet{2007ApJ...664..481B} together with the sample of SWIRE/SDSS QSO from \citet{2008MNRAS.386.1252H}. The very low mass member population (with or without mid-IR excess associated to the presence of a circumstellar disc) and the QSO populations occupy two very distinct areas of the diagram.  We note that 97.3\% of the confirmed very low mass stellar and substellar members of \citet{2007ApJ...664..481B} and only 2.3\% of the SWIRE/SDSS quasars fall within the following area:
\begin{eqnarray}
{i \rm - J} \ge 1.12 \times ({\rm J}-[3.6]) - 1.0 \label{sel1}
\end{eqnarray}

and we use this condition to discard 1 extragalactic contaminant in our catalogue. Another source has a J-[3.6] color much bluer than any of the confirmed members and is discarded as well. A total of 16 sources remain after this first selection. 

\begin{figure}
   \centering
   \includegraphics[width=0.45\textwidth]{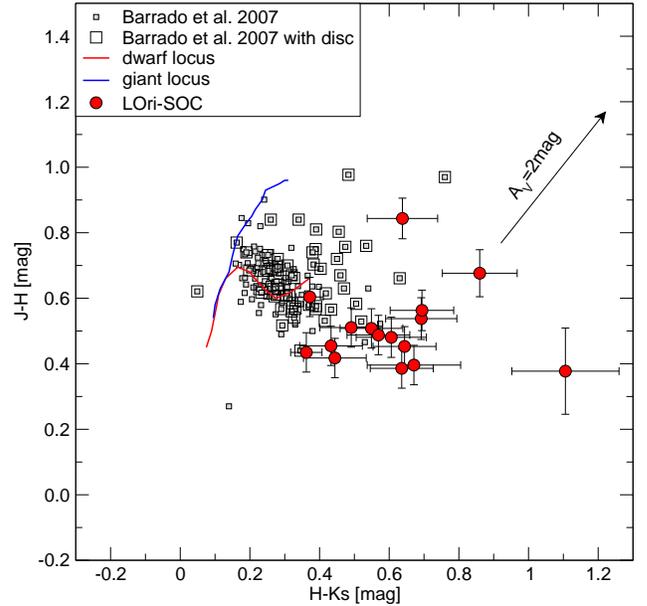}
      \caption{J-H vs H-Ks color-color diagram of the LOri-SOC candidates (red dots) and confirmed members from \citet{2007ApJ...664..481B} (grey squares, and an additional large open square for objects of their sample with mid-IR excess associated to a disc). The giant locus is represented with a thick blue curve and the dwarf locus by a thick red curve. An A$_{\rm V}$=2~mag reddening vector is represented.  \label{jh_hk_soc}}
   \end{figure}

A careful look at a J-H vs H-Ks color-color diagram (Fig.~\ref{jh_hk_soc}) shows that all 16 previously selected sources (including the one without [3.6] photometry) have colors consistent with dwarfs from the association, and we select them as cluster member candidates. Their astrometry and photometry is reported in Table~\ref{table_soc}.

As we use part of the same CFHT12K dataset, the current survey overlaps with that of \citet{2004ApJ...610.1064B} and 14 of their sources fall in the common area. Seven of them are too bright and saturated in our new near-infrared images, and we verify that the remaining seven confirmed members are successfully selected as candidate members in our independent analysis, providing a sanity check of our selection strategy. A total of 9 new candidate members are found by our analysis. \citet{2004ApJ...610.1064B} estimate a contamination rate of 25\% within their sample selected based on optical and near-infrared luminosities and colours. Our selection includes optical, near- and mid-infrared colors, and the contamination rate is expected to be similar or lower. 

\subsection{MAD catalogue}
\begin{figure}
   \centering
   \includegraphics[width=0.45\textwidth]{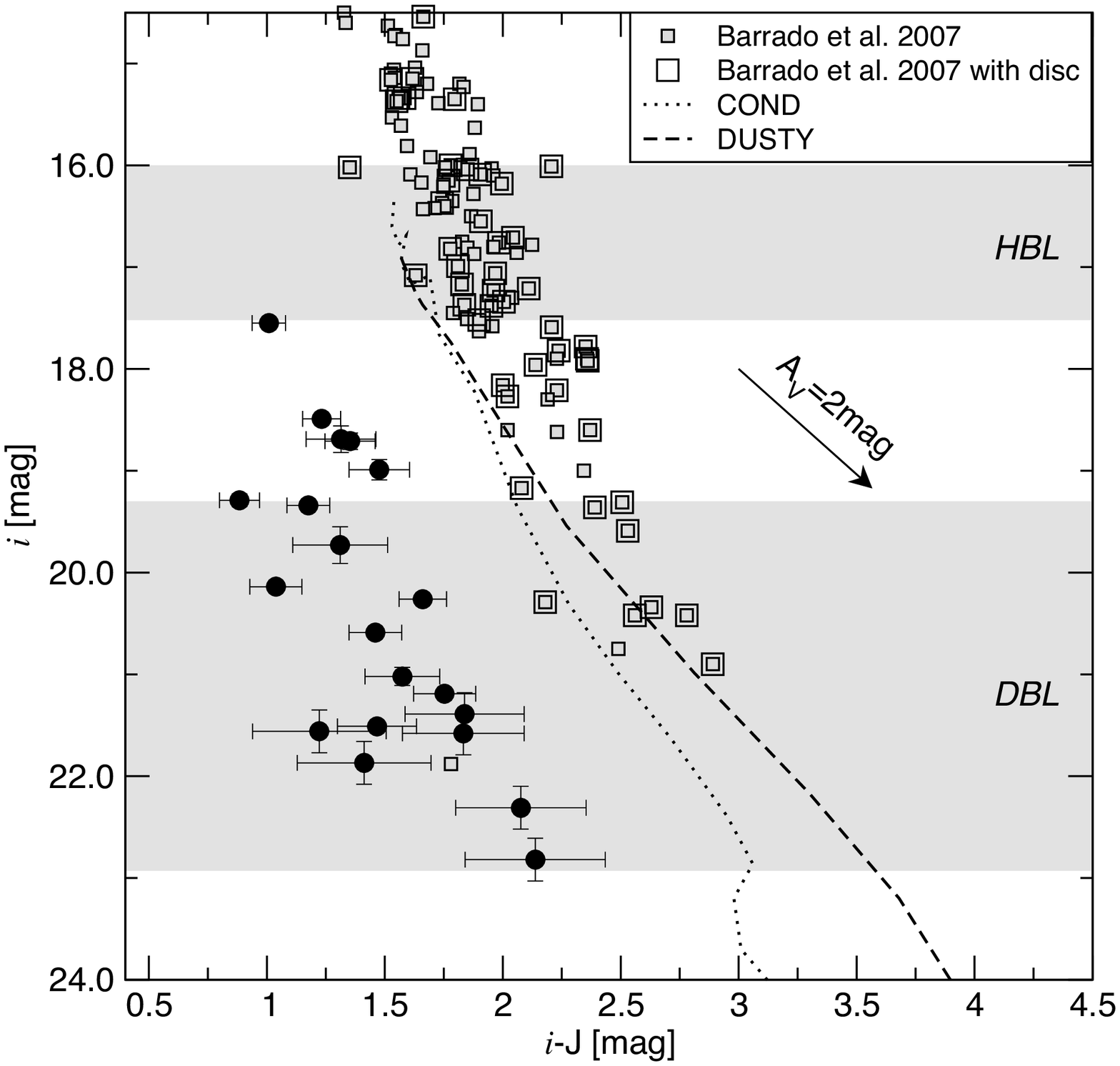}
      \caption{Same as Fig.~\ref{i_ij_soc} for the LOri-MAD sample (black dots).  \label{i_ij_mad}}
   \end{figure}

A total of 20 out of 32 sources detected in the MAD images have an optical counterpart in either the CFHT12K or Suprime-Cam $i$-band images, as reported in Table~\ref{table_mad}. We use a $i$ vs $i$-J diagram to select possible candidate members among these 20 sources. Figure~\ref{i_ij_mad} shows that none of the 20 sources has colors consistent with the cluster's isochrone. Most of them are therefore likely foreground or background coincidences and discarded for the rest of the analysis. The selection among the remaining 12 sources is done case-by-case using the near-infrared photometry.

\begin{figure}
   \centering
   \includegraphics[width=0.45\textwidth]{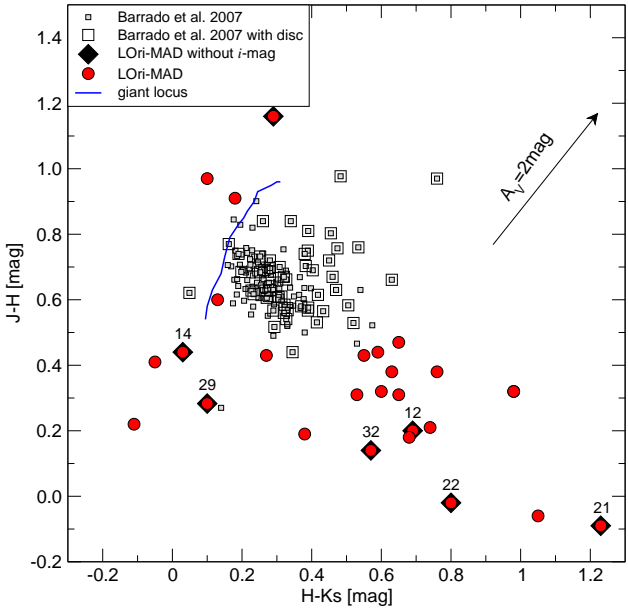}
      \caption{Same as Fig.~\ref{jh_hk_soc} for the LOri-MAD sample (red dots). Sources whose membership could not be assessed  in Fig.~\ref{i_ij_mad} because of the lack of $i$-band photometry are overplotted with black diamond. Their LOri-MAD number is given. \label{jh_hk_mad}}
  \end{figure}

\begin{figure}
   \centering
   \includegraphics[width=0.45\textwidth]{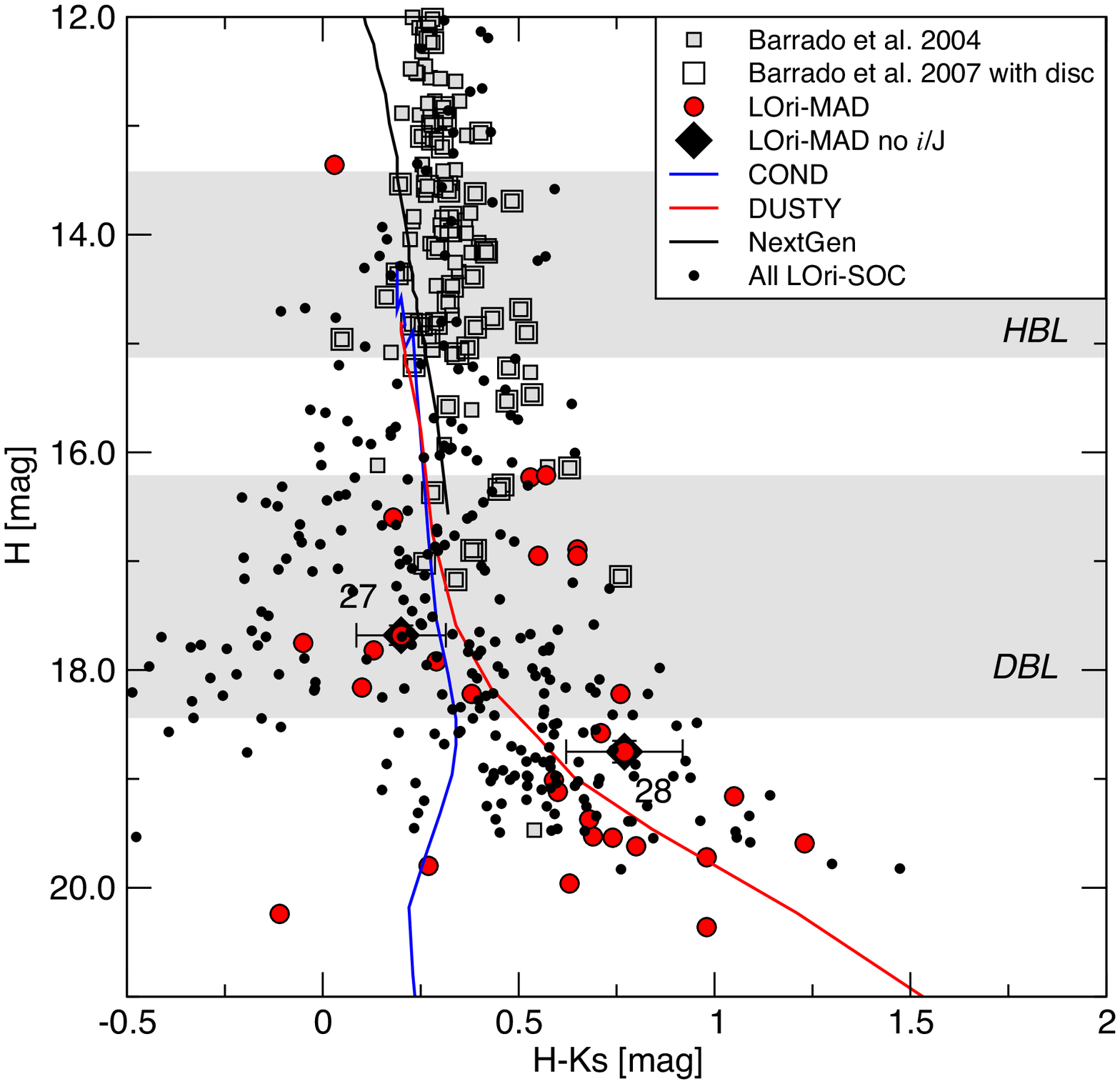}
      \caption{H vs H-Ks color-magnitude diagram for the LOri-MAD sample (red dots). The 2 sources whose membership could not be assessed  in Fig.~\ref{i_ij_mad} and \ref{jh_hk_mad} because of the lack of $i$ and J-band photometry are overplotted with a black diamond. Their LOri-MAD number is given. The sources detected in the SOC images are overplotted with black dots, and shows that background objects and cluster members cover the same colour range. The membership of LOri-MAD-27 and -28 thus remains undetermined. The hydrogen (HBL) and deuterium burning limits (DBL) according to the DUSTY models for ages between 1--5~Myr and distances in the range 325--450~pc are  represented with grey shaded areas \label{h_hk_mad}}
  \end{figure}

LOri-MAD-29 is the bright HD~36861C, the F8V companion to the massive $\lambda-$Ori \citep{1985A&AS...60..183L}. Figure~\ref{jh_hk_mad} shows that LOri-MAD-14 (2MASS~J05350590+0955533) and LOri-MAD-31 have near-infrared colors indicative of a giant status in a J-H vs H-Ks color-color diagram and are discarded for the rest of the analysis. LOri-MAD-19, -24 and -30 are detected only in one band and their membership cannot be tested and remains undetermined. LOri-MAD-12, -21, -22 and -32 have near-infrared colors inconsistent with the cluster sequence in a J-H vs H-Ks color-color diagram and are most likely extragalactic contaminants. LOri-MAD-27 and -28 only have MAD H and Ks photometry. Fig.~\ref{h_hk_mad} shows that they both have luminosities and a color consistent with the cluster sequence within the uncertainties. Field dwarfs unfortunately have H-Ks colours indistinguishable from young cluster members, and the nature and membership of LOri-MAD -27 and -28 is therefore undetermined. Taking into account that the other LOri-MAD sources with full photometry all turn out to be background/foreground objects, these 2 sources have a high probability of being contaminants as well. Table~\ref{table_mad} gives a summary of the membership analysis. 

We note that within our sensitivity limits, a 10\,000~light year diameter galaxy would have been resolved by our MCAO images as far as $z\approx$1.5. Finally, six LOri-MAD sources have a {\it Spitzer} IRAC counterpart. The corresponding photometry is reported in Table~\ref{table_spitzer} for information, but is not discussed further since all six sources are discarded from the sample of member candidates.

\section{Visual multiple systems}

With an average resolution of $\approx$0\farcs1, the MAD images resolve a number of visual multiple systems. \\

\noindent\emph{LOri-MAD-20 \& -19:} the two sources form a visual pair with a separation of 0\farcs515$\pm$0\farcs010 and a position angle of P.A=72.4\degr. The relative astrometry is quite uncertain as these two objects suffer from the above mentioned elongation/doubling of the PSF. The primary is much bluer than the cluster sequence in all color-magnitude and color-color diagrams, suggesting that it does not belong to the association. The secondary is not detected in the Ks image and no color information is available.\\

\noindent\emph{LOri-MAD-23 \& -22:} form a visual pair with a separation of 0\farcs944$\pm$0\farcs004 and a position angle of P.A=142.9\degr. It is detected and resolved in the SofI J-band image, while only the primary is detected in the optical CFHT12K and Subaru images. Spectroscopy as well as proper motion measurements are required to confirm the nature of this pair, but the colors and luminosities of the two components are not consistent with the cluster sequence.\\

\noindent\emph{LOri-MAD-30 \& -29 (HD~36861C):} is resolved for the first time in the MAD Ks image and form a visual pair with a separation of 0\farcs384$\pm$0.004 and P.A=213.2\degr. \citet{1985A&AS...60..183L} classified LOri-MAD-29 as a F8V companion to the $\lambda-$Ori~AB pair (separation 28\farcs6). The membership of LOri-MAD-30 is undetermined since it is detected in the Ks-band only. If physical, the pair HD~36861Cab would be made of an F8V star and a $\approx$0.040~M$_{\sun}$ brown dwarf (according to the DUSTY models and assuming a distance of 400~pc at an age of 5~Myr) with a projected separation of $\approx$155~AU. The paucity of brown dwarfs observed at such separation around solar type or intermediate type stars in the field \citep{2007lyot.confE..31M} or in other young associations \citep{2007A&A...464..581K, 2008ApJ...679..762K} make the potenial existence of such substellar companion in the core of the young $\lambda-$Orionis cluster an interesting case.

\section{Conclusions}
We present a new analysis combining optical and near-infrared seeing limited images as well as diffraction limited adaptive optics images of the inner 5\farcm3 of the $\lambda-$Orionis cluster. We report the discovery of 9 new very low mass and substellar candidate members with 13.27$\le$Ks$\le$17.25~mag, and independently report 7 previously known members. Spectroscopy is required to confirm the nature and membership to the association of the 9 new candidate members. The high contrast images allow us to rule out the presence of substellar companion (M$\ge$0.07~M$_{\sun}$) in the range 1--4\arcsec around the massive $\lambda-$Ori O8III star.

\begin{acknowledgements}
The authors are grateful to Paola Amico for her support at ESO. We thank A. Alonso-Herrero for her help and suggestions. We thank our anonymous referee for a prompt and constructive review that helped improve this article. H. Bouy acknowledges the funding from the European Commission's Sixth Framework Program as a Marie Curie Outgoing International Fellow (MOIF-CT-2005-8389). N. Hu\'elamo and D. Barrado y Navascu\'es are funded by Spanish grants MEC/ESP2007-65475-C02-02, MEC/Consolider-CSD2006-0070 and CAM/PRICIT-S-0505/ESP/0361.

This work is based on observations obtained with the MCAO Demonstrator (MAD) at the VLT (ESO Public data release), which is operated by the European Southern Observatory. The MAD project is led and developed by ESO with the collaboration of the INAF-Osservatorio Astronomico di Padova (INAF-OAPD) and the Faculdade de Ci\^encias de Universidade de Lisboa (FCUL). Based on observations made with ESO Telescopes at the La Silla or Paranal Observatories under programmes 082.C-0724, 080.D-0532, 67.C-0042, 074.C-0084, and 074.C-0628. Based in part on data collected at Subaru Telescope, which is operated by the National Astronomical Observatory of Japan. This work is based in part on observations made with the Spitzer Space Telescope, which is operated by the Jet Propulsion Laboratory, California Institute of Technology under a contract with NASA. This publication makes use of data products from the Two Micron All Sky Survey, which is a joint project of the University of Massachusetts and the Infrared Processing and Analysis Center/California Institute of Technology, funded by the National Aeronautics and Space Administration and the National Science Foundation. This work has made use of the Vizier Service provided by the Centre de Donn\'ees Astronomiques de Strasbourg, France \citep{Vizier}. This research used the facilities of the Canadian Astronomy Data Centre operated by the National Research Council of Canada with the support of the Canadian Space Agency. 

\end{acknowledgements}

\bibliographystyle{aa}

\input{table_soc}

\input{table_mad3}

\input{table_spitzer}

\end{document}

%% file: table_soc.tex
\begin{center} 
\begin{table*}
\scriptsize
\caption{$\lambda$-Ori candidates detected in the SofI-Omega2000-CFHT12K-SuprimeCam (SOC) images \label{table_soc}}
\begin{tabular}{lccccccccccc}
\hline\hline
Name & RA         & Dec       & $i$ & J & H & Ks & [3.6] & [4.5] & [5.8] & [8.0]  & LOri-CFHT \\
     & (J2000)    & (J2000)   & [mag] & [mag]& [mag]& [mag]& [mag]& [mag]& [mag]& [mag] \\
\hline
LOri-SOC-1	 &  05:34:55.53 & +09:56:10.09 & 16.39$\pm$0.01 & 14.59$\pm$0.06 & 14.16$\pm$0.01 & 13.80$\pm$0.05 & 13.44$\pm$0.01 & 13.85$\pm$0.02 & 13.37$\pm$0.03 & \nodata     \\
LOri-SOC-2	 &  05:34:56.67 & +09:54:54.12 & 16.21$\pm$0.01 & 14.27$\pm$0.06 & 13.67$\pm$0.01 & 13.30$\pm$0.04 & 13.34$\pm$0.01 & 12.90$\pm$0.01 & 13.20$\pm$0.03 & 13.06$\pm$0.05 &   \\
LOri-SOC-3	 &  05:35:00.90 & +09:54:40.39 & 20.84$\pm$0.02 & 18.12$\pm$0.06 & 17.58$\pm$0.02 & 16.89$\pm$0.10 & 17.13$\pm$0.05 &  \nodata       &  \nodata       & \nodata    \\
LOri-SOC-4	 &  05:35:00.95 & +09:58:20.33 & 18.34$\pm$0.01 & 16.02$\pm$0.06 & 15.62$\pm$0.01 & 14.95$\pm$0.13 & 14.77$\pm$0.02 & 14.64$\pm$0.02 & 14.07$\pm$0.05 & 13.68$\pm$0.07 & 143 \\
LOri-SOC-5	 &  05:35:03.03 & +09:55:47.31 & 22.08$\pm$0.04 & 18.73$\pm$0.06 & 18.36$\pm$0.12 & 17.25$\pm$0.10 &  \nodata       & 16.67$\pm$0.07 & 15.65$\pm$0.22 & \nodata     \\
LOri-SOC-6	 &  05:35:03.73 & +09:54:13.93 & 17.33$\pm$0.01 & 15.65$\pm$0.06 & 15.14$\pm$0.01 & 14.65$\pm$0.09 & 14.43$\pm$0.02 & 14.24$\pm$0.02 & 14.37$\pm$0.07 & 13.83$\pm$0.08 &  \\
LOri-SOC-7	 &  05:35:04.46 & +09:57:32.08 & 20.45$\pm$0.01 & 17.42$\pm$0.06 & 16.94$\pm$0.01 & 16.33$\pm$0.10 & 16.03$\pm$0.02 & 16.42$\pm$0.05 &  \nodata       & \nodata  & 162   \\
LOri-SOC-8	 &  05:35:06.32 & +09:58:01.38 & 17.61$\pm$0.01 & 15.59$\pm$0.06 & 15.17$\pm$0.01 & 14.73$\pm$0.09 & 14.22$\pm$0.01 & 14.07$\pm$0.02 & 14.00$\pm$0.06 & \nodata & 128    \\
LOri-SOC-9	 &  05:35:07.08 & +09:54:01.51 & 16.94$\pm$0.01 & 14.75$\pm$0.06 & 14.24$\pm$0.01 & 13.69$\pm$0.09 & 13.63$\pm$0.01 & 12.58$\pm$0.01 & 12.60$\pm$0.02 & 11.78$\pm$0.02 & 104 \\
LOri-SOC-10	 &  05:35:07.79 & +09:55:21.84 & 19.80$\pm$0.01 & 16.89$\pm$0.06 & 16.32$\pm$0.01 & 15.63$\pm$0.09 & 14.80$\pm$0.02 & 14.30$\pm$0.02 &  \nodata       & \nodata     \\
LOri-SOC-11	 &  05:35:09.10 & +09:54:36.07 & 19.00$\pm$0.01 & 16.46$\pm$0.06 & 16.00$\pm$0.01 & 15.36$\pm$0.09 & 14.91$\pm$0.02 & 14.99$\pm$0.02 & 14.84$\pm$0.10 & \nodata     \\
LOri-SOC-12	 &  05:35:10.65 & +09:57:23.31 & 21.62$\pm$0.03 & 18.66$\pm$0.06 & 17.98$\pm$0.04 & 17.12$\pm$0.10 & 16.78$\pm$0.04 & 16.06$\pm$0.04 &  \nodata       & \nodata     \\
LOri-SOC-13	 &  05:35:10.78 & +09:56:06.48 & 18.16$\pm$0.01 & 15.94$\pm$0.06 & 15.56$\pm$0.01 & 14.92$\pm$0.09 & 14.30$\pm$0.01 & 13.59$\pm$0.01 &  \nodata       & 13.77$\pm$0.08 &   \\
LOri-SOC-14	 &  05:35:10.97 & +09:57:43.95 & 16.46$\pm$0.01 & 14.16$\pm$0.06 & 13.70$\pm$0.01 & 13.27$\pm$0.09 & 13.01$\pm$0.01 & 12.68$\pm$0.01 & 12.70$\pm$0.02 & 12.58$\pm$0.03 & 76 \\
LOri-SOC-15	 &  05:35:11.14 & +09:57:19.68 & 16.60$\pm$0.01 & 14.69$\pm$0.06 & 14.20$\pm$0.01 & 13.63$\pm$0.09 & 13.14$\pm$0.01 & 12.71$\pm$0.01 & 12.63$\pm$0.02 & 11.99$\pm$0.02 & 96 \\
LOri-SOC-16	 &  05:35:14.17 & +09:54:08.05 & 20.92$\pm$0.02 & 18.04$\pm$0.06 & 17.20$\pm$0.01 & 16.56$\pm$0.10 & 16.06$\pm$0.02 &  \nodata       &  \nodata       & \nodata    & 167  \\
\hline
\end{tabular}

{\bf Note -- Uncertainties correspond to the measurement uncertainties. Zeropoint uncertaintes as given in the text should be added quadratically. }
\end{table*}
\end{center}

%% file: table_mad3.tex
\begin{center} 
\begin{table*}
\scriptsize
\caption{Catalogue of sources detected in the MAD images \label{table_mad}}
\begin{tabular}{lcccccccccl}\hline\hline
LOri-MAD & RA      & Dec         & $i$              & J              & H              & Ks             &  M1   &  M2   &   Comment \\
         & (J2000) & (J2000)     & [mag]            & [mag]          & [mag]          & [mag]          &       &       &      \\
\hline
1  & 05:35:04.264 & +09:55:50.61 & 20.14$\pm$0.01   & 19.10$\pm$0.06 & 19.16$\pm$0.12 & 18.11$\pm$0.13 & NM  &  &      \\
2  & 05:35:05.213 & +09:56:10.77 & 20.26$\pm$0.01   & 18.60$\pm$0.06 & 18.22$\pm$0.08 & 17.46$\pm$0.08 & NM  &  & PSF anomaly       \\
3  & 05:35:05.309 & +09:56:27.38 & 21.51$\pm$0.05   & 20.04$\pm$0.06 & 19.72$\pm$0.14 & 18.74$\pm$0.16 & NM  &  & PSF anomaly       \\
4  & 05:35:05.338 & +09:56:15.25 & 18.69$\pm$0.13   & 17.38$\pm$0.06 & 16.95$\pm$0.06 & 16.40$\pm$0.05 & NM  &  & PSF anomaly       \\
5  & 05:35:05.432 & +09:56:30.24 & 21.02$\pm$0.09   & 19.46$\pm$0.06 & 19.01$\pm$0.10 & 18.42$\pm$0.14 & NM  &  & PSF anomaly       \\
6  & 05:35:05.454 & +09:55:32.61 & 19.73$\pm$0.18   & 18.43$\pm$0.06 & 17.82$\pm$0.09 & 17.69$\pm$0.08 & NM  &  &     \\
7  & 05:35:05.531 & +09:56:11.96 & 21.56$\pm$0.21   & 20.34$\pm$0.07 & 19.96$\pm$0.15 & 19.33$\pm$0.21 & NM  &  &     \\
8  & 05:35:05.576 & +09:55:26.06 & 20.59$\pm$0.02   & 19.13$\pm$0.06 & 18.59$\pm$0.12 & 18.06$\pm$0.13 & NM  &  &     \\
9  & 05:35:05.614 & +09:56:30.69 & 22.82$\pm$0.21   & 20.68$\pm$0.07 & 20.36$\pm$0.21 & 19.38$\pm$0.26 & NM  &  & PSF anomaly       \\
10 & 05:35:05.665 & +09:56:13.59 & 21.87$\pm$0.21   & 20.47$\pm$0.07 & 20.24$\pm$0.17 & 20.35$\pm$0.46 & NM  &  &     \\
11 & 05:35:05.691 & +09:55:48.15 & 22.31$\pm$0.21   & 21.78$\pm$0.14 & 19.80$\pm$0.14 & 19.53$\pm$0.25 & NM  &  &     \\
12 & 05:35:05.716 & +09:56:35.36 & \nodata          & 19.74$\pm$0.06 & 19.53$\pm$0.13 & 18.84$\pm$0.18 &     & NM  &      \\
13 & 05:35:05.889 & +09:56:23.16 & 21.19$\pm$0.02   & 19.45$\pm$0.06 & 19.12$\pm$0.10 & 18.52$\pm$0.13 & NM  &  & PSF anomaly       \\
14 & 05:35:05.901 & +09:55:53.35 & \nodata          & 13.80$\pm$0.06 & 13.36$\pm$0.05 & 13.33$\pm$0.04 &     & NM  &      \\
15 & 05:35:05.949 & +09:55:47.82 & 21.39$\pm$0.21   & 19.55$\pm$0.06 & 19.37$\pm$0.11 & 18.69$\pm$0.15 & NM  &  &     \\
16 & 05:35:06.001 & +09:55:39.33 & 18.71$\pm$0.08   & 17.36$\pm$0.06 & 16.89$\pm$0.06 & 16.24$\pm$0.05 & NM  &  &     \\
17 & 05:35:06.085 & +09:55:48.51 & 21.58$\pm$0.21   & 19.76$\pm$0.06 & 19.54$\pm$0.12 & 18.80$\pm$0.16 & NM  &  &     \\
18 & 05:35:06.243 & +09:56:34.68 & 19.29$\pm$0.06   & 18.42$\pm$0.06 & 18.22$\pm$0.08 & 17.84$\pm$0.10 & NM  &  & PSF anomaly       \\
19 & 05:35:06.503 & +09:56:32.60 & \nodata          & \nodata        & 18.58$\pm$0.09 & $>$19.1   0.11 &     &  & Bin. B, PSF anomaly  \\
20 & 05:35:06.536 & +09:56:32.45 & 18.99$\pm$0.1    & 17.51$\pm$0.06 & 16.60$\pm$0.04 & 16.42$\pm$0.08 & NM  &  & Bin. A, PSF anomaly  \\
21 & 05:35:06.731 & +09:56:19.68 & \nodata          & 19.50$\pm$0.06 & 19.59$\pm$0.12 & 18.36$\pm$0.12 &     & NM  &      \\
22 & 05:35:06.757 & +09:55:46.82 & \nodata          & 19.60$\pm$0.06 & 19.62$\pm$0.13 & 18.82$\pm$0.15 &     & NM  & Bin. B     \\
23 & 05:35:06.794 & +09:55:47.60 & 18.49$\pm$0.01   & 17.26$\pm$0.06 & 16.95$\pm$0.06 & 16.30$\pm$0.05 & NM  &  & Bin. A    \\
24 & 05:35:07.577 & +09:55:37.36 & \nodata          & $>$20.1        & 20.68$\pm$0.16 & $>$19.6        &     &  &      \\
25 & 05:35:07.672 & +09:55:48.02 & 17.55$\pm$0.01   & 16.55$\pm$0.06 & 16.23$\pm$0.05 & 15.70$\pm$0.05 & NM  &  &     \\
26 & 05:35:07.811 & +09:55:30.79 & 19.34$\pm$0.01   & 18.17$\pm$0.06 & 17.75$\pm$0.09 & 17.80$\pm$0.10 & NM  &  &     \\
27 & 05:35:07.956 & +09:56:08.77 & \nodata          & $>$15.1        & 17.68$\pm$0.09 & 17.48$\pm$0.07 &     &  & membership undertermined \\
28 & 05:35:07.979 & +09:55:52.85 & \nodata          & $>$17.6        & 18.75$\pm$0.10 & 17.98$\pm$0.11 &     &  & membership undertermined \\
29 & 05:35:08.160 & +09:55:34.38 & \nodata          & 9.39$\pm$0.03  &  9.11$\pm$0.03 &  9.01$\pm$0.02 &     & & Bin. A, F8V, Member \\
30 & 05:35:08.138 & +09:55:34.27 & \nodata          & \nodata        & $>$13.3        & 14.75$\pm$0.05 &     & & Bin B., membership undertermined     \\
31 & 05:35:08.245 & +09:55:41.24 & \nodata          & 19.08$\pm$0.06 & 17.92$\pm$0.09 & 17.63$\pm$0.09 &     & NM  &     \\
32 & 05:35:08.443 & +09:55:44.62 & \nodata          & 16.36$\pm$0.06 & 16.21$\pm$0.06 & 15.64$\pm$0.05 &     & NM  &     \\
\hline
\end{tabular}

Note -- M1: membership assessed from the location in a $i$ vs $i$-J color-magnitude diagram only. M2: membership assessed from the location in a J-H vs H-Ks color-color diagram only. NM=not member. Uncertainties correspond to the measurement uncertainties. Zeropoint uncertainties as given in the text should be added quadratically.
\end{table*}
\end{center}

%% file: table_spitzer.tex
\begin{table*}
\caption{Spitzer IRAC photometry (mag) of LOri-MAD sources \label{table_spitzer}}
\begin{tabular}{lcccccc}\hline\hline
Name        & 3.6~$\mu$m  & 4.5~$\mu$m &5.8~$\mu$m & 8.0~$\mu$m \\
\hline
LOri-MAD-1 & 17.67$\pm$0.08 & \nodata & \nodata & \nodata \\
LOri-MAD-4 & 15.84$\pm$0.02 & 16.00$\pm$0.03 & \nodata & \nodata \\
LOri-MAD-6 & 17.64$\pm$0.07 & \nodata & \nodata & \nodata \\
LOri-MAD-8 & 17.42$\pm$0.03 & \nodata & \nodata & \nodata \\
LOri-MAD-14 & 13.42$\pm$0.02 & 13.09$\pm$0.01 & 13.26$\pm$0.03 & 13.83$\pm$0.08 \\
LOri-MAD-16 & \nodata & 16.83$\pm$0.07 & \nodata & \nodata \\
\hline
\end{tabular}

Note -- Uncertainties correspond to the measurement uncertainties. Zeropoint uncertaintes as given in the text should be added quadratically.
\end{table*}